\documentclass[12pt]{article}
\begin{document}
\begin{flushright}
IITAP-97-011 \\
hep-ph/9709303 \\
September 1997 \\                
\end{flushright}
\vspace*{1.cm}
\begin{center}
{ \large \bf A New Qualitative Prediction of the Parton Model \\
for High Energy Hadron Collisions} \\
\vspace*{0.3cm}
{\large  Victor T. Kim${}^{\dagger}$, Grigorii B. Pivovarov${}^{\ddagger}$ 
and James P. Vary ${}^{\S}$ }\\
\vspace*{0.3cm}
{\em ${}^\dagger$ :
St.Petersburg Nuclear Physics Institute,
 188350 Gatchina,
Russia} {\footnotemark} \\
{\em ${}^\ddagger$ :
Institute for Nuclear
Research, 117312 Moscow, Russia} {\footnotemark} \\
{\em $\S $ : International Institute of Theoretical and Applied Physics,\\
     Iowa State University, Ames, Iowa 50011-3022, USA} {\footnotemark}
\end{center}
\vspace{0.5cm}
\begin{center}
{\large \bf Abstract}
\end{center}
Inclusive single jet production in hadron collisions is 
considered. It is shown that the QCD parton model predicts   
a nonmonotonic dependence of the inclusive cross section on
the fraction of the energy deposited in the jet registered,
if it is normalized on the same cross section measured at 
another collision energy. Specifically, if the cross section 
is normalized by the one measured at a higher collision 
energy, it possesses a minimum which depends on jet rapidity. 
This prediction can be tested at the Fermilab Tevatron, at the 
CERN LHC, and at the Very Large Hadron Collider under discussion.

\vspace*{4cm}

to appear in {\it Foundation of Physcics}

\addtocounter{footnote}{-2}
\footnotetext{e-mail: $kim@pnpi.spb.ru$}
\addtocounter{footnote}{1}
\footnotetext{e-mail: $gbpivo@ms2.inr.ac.ru$}
\addtocounter{footnote}{1}
\footnotetext{e-mail: $jvary@iastate.edu$}

\newpage

The parton model, improved with the QCD running of the coupling and of
the structure functions, has the well-deserved status of a paradigm.
Its diverse justifications are of two kinds: the qualitative justifications
of the early days of the parton model (like the prediction of the 
relation between structure functions $F_1$ and $F_2$ governed by the spin
of the partons (Gross relation), for a review see, e.g., 
Ref. \cite{Iof84}) and the present-day
abundant quantitative justifications (for a recent review, including
a list of structure function parameterizations, see Ref.
\cite{CTEQ95}).

The advent of the Fermilab Tevatron and the CERN LHC 
provides a new testing ground for the parton
model---the kinematic conditions when the energies of the produced hadrons
are large enough to be described by
perturbation theory and,  at the same time, are much smaller than the total
energy of the collision (semi-hard kinematics). 
Because the parton model was originally invented and subsequently tested for
the hard kinematics, the second condition makes it plausible
that a substantial modification of the parton model will be needed to describe
this semi-hard kinematic region. Note that for hard kinematics the
transverse energy of the produced hadrons is on the order of the
total collision energy. In particular, some fusion of Regge theory (see,
e.g., Ref. \cite{Col82})  and perturbative QCD may be required for the
semi-hard kinematics. The  Balitsky-Fadin-Kuraev-Lipatov (BFKL)
equation \cite{Lip76} and, more generally, the experience with the
resummation of the large energy logarithms
(for a review see Ref. \cite{Lip97}) may be useful in
constructing this new model. 

The range of applicability of the QCD improved parton model
is a subject of controversy at the moment. There are statements
(see, e.g., \cite{Bal97,Ynd96}) that the fitting capacity 
of the conventional parton model is sufficient to accomodate all the 
data on parton structure functions available at the 
semi-hard kinematics. On the other
hand, the data from HERA on forward jet production at small-x
\cite{HERA98} may be interpreted as a manifestation of the BFKL 
Pomeron \cite{Bar96}, which seems to be beyond the limits 
of the conventional parton model. 
The situation is further complicated by the observation that
the range of applicability of the parton model may be different 
for different observables. In particular, the cross sections of 
processes with specific kinematics exhibit
breakdown of the applicability of finite order
perturbative QCD via the loss of insensitivity to the choice of 
the normalization scale. On the other hand, some dedicated combinations 
(ratios) of cross sections may be less sensitive to the inclusion of the
higher order corrections. An example is the scaled cross section 
ratio \cite{UA285,CDF93,CDF96,Hus98} since, as it 
follows from  Ref. \cite{Ell93}, it is relatively 
insensitive to the inclusion of the next-to-leading (NLO) correction.

Under these circumstances, it is crucial to have {\it qualitative}
predictions from the conventional QCD-improved parton model (without
resummation of the energy logarithms) for the new kinematic domain. If the
predictions would turn out qualitatively incorrect, a substitute for the
parton model would become indispensable.

In this paper, we present such a prediction. It is a prediction
for the ratio of inclusive single jet production at a smaller energy 
$\sqrt{s_N}$
of the hadron collision to the one at a higher energy $\sqrt{s_D}$:
\begin{equation}
\label{rdef}
R(x, y) = \bigg( \frac{s_N d\sigma(s_N)}{dxdy} \bigg) \Big/ 
\bigg( \frac{s_D d\sigma(s_D)}{dxdy} \bigg) .
\end{equation}
Here the cross section is made dimensionless by the rescaling with the 
corresponding total invariant energy of the collision squared
$s_N (s_D)$. 
The ratio depends on the (pseudo)rapidity $y=1 / 2 \ln (k_+ / k_-)$,
where $k_\pm = E \pm k_3$ are 
the light-cone components  of the momentum of the produced jet, and on the
fraction of the energy 
$x = (k_+ + k_-) / \sqrt{s_i}$, $i=N,D$ deposited in the jet produced 
($s_N$ is used for the definition of $x$ in the numerator, $s_D$ in the
denominator, so $x$ varies from zero to unity for both energies).
Note that this scaling variable coincides in the center of mass system
with 
$x_R = E/E_{max} = 2 E/ \sqrt{s}$ , 
the radial Feynman variable.

The ratio $R$ (taken at $y \simeq 0$, i.e., for jets perpendicular 
to the collision axes) was used in Refs. \cite{UA285,CDF93,CDF96} 
as a means to test QCD predictions
for scaling violations. Note that without scaling violations the
ratio
$R$ is {\it exactly} unity. The $x$-dependence of $R$ comes from the
presence
of 
$\Lambda_{QCD}$ in the running coupling and in the parton distribution 
functions.

Perturbative QCD calculations of Ref. \cite{Ell93} with hard kinematics 
($Q^2  \sim  s $) predict for $R$ at $y=0$ 
a steep increase around the value of 1.8-1.9 for $x$ growing in the 
range above 0.1  (for the case $\sqrt{s_N}$/$\sqrt{s_D}$ =
0.63 TeV/1.8 TeV ).   For moderate $x$, the prediction is in reasonable
agreement with CDF data  \cite{CDF93,CDF96}.
For $x<0.1$ calculations are above the preliminary data of CDF \cite{CDF96}.
This was among the reasons for the conclusion of Refs. \cite{Ell93,Gie96} that 
NLO perturbative QCD \cite{Ell90} with hard kinematics 
is insufficient for the description of absolute cross section
of jets with  transverse energy less  than 50 GeV within 
accuracy $10 \% $. It was shown in Ref. \cite{KP96b} that 
resummation of the energy 
logarithms restores the agreement between theory and experiment.

In this paper we present the following result: 
the QCD-improved parton model predicts that $R$ is not a monotonic
function of its arguments, i.e. the single jet production cross section, if
measured in the natural units of the same cross section taken at another
(higher) energy of the collision, has extrema. Namely, it has minima 
(``dips''): 
there is a value of $x$ for each $y$ with the smallest ratio of jets
produced. The reason this fact was overlooked is that for $y=0$ 
(the only value for which
the calculations were reported earlier) the minimum is at a
value of $x$ too small to be inside the acceptance of the existing detectors
($x_{dip}(y=0)<0.01$ at the Tevatron). 

Fig. 1 presents the ratio for energies 0.63 TeV/1.8 TeV at the 
Fermilab Tevatron, Fig. 2 for energies 6 TeV/14 TeV at the CERN LHC,
and Fig. 3 for energies 6 TeV/100 TeV of the Very Large Hadron Collider
(VLHC) proposal \cite{VLHC}. 
Each curve on the plots presents the dependence of the ratio 
on $x$ at different values of rapidity $y$.
Each curve ends at a lowest value of $x$ where
$\alpha_S(Q^2) = 4 \pi / \big[ (11 - 2/3 n_f) 
\ln(Q^2/\Lambda_{QCD}^2) \big] $ 
has the value of about 0.5 (it corresponds
to $Q=0.7$ GeV, and $Q$ was taken to be half of the transverse
energy of the jet produced).
For lower values of $x$ perturbative theory becomes unreliable 
because the coupling approaches unity. 

There is another lower bound on the values of $x$ at which our plots
make sense, because there is a lowest energy for which the jet may 
be resolved. This energy is accepted now to be around 5 GeV 
\footnote{At the  HERA, lepton-hadron collider, jets are resolved 
from $E_\perp=3$ GeV, and at the Tevatron, hadron-hadron collider,
e.g., CDF Collaboration used to tagging jets from $E_\perp=8$ GeV 
\cite{CDF98}. }
which corresponds to $x>0.016$ for the Tevatron (Fig. 1), 
and to $x>0.0017$ for the LHC (Fig. 2).

The cross sections for the plots of Figs. 1-3 were calculated 
with the formula
\begin{equation}
\label{cross}
\frac{sd\sigma}{dydx}=\frac{\pi \big[ C_A\alpha_S(Q^2) \big]^2}{x}
\big[ I(x,y,Q^2)+I(x,-y,Q^2) \big],
\end{equation}
where $C_A=3$ for $SU(3)$, and the two terms in the brackets of the
right-hand-side
correspond to the contributions with balancing jet of rapidity over 
($I(x,y,Q^2)$) and below ($I(x,-y,Q^2$)) the one of the registered jet 
(the balancing jet is the unregistered jet whose transverse
momentum balances the one of the registered jet; 
the presence of two terms corresponds to the symmetry
of the cross section by which it is an even function of $y$);
$I(x,y,Q^2)$ is a convolution of the parton distribution functions;
the normalization point for the coupling and for the parton distribution
functions is chosen to be $Q = 0.5 E_{\perp}$, where $E_\perp$ is the 
transverse energy of the jet produced \cite{CTEQ95}.

\begin{figure}[htb]
\vskip 7.5 cm
\includegraphics{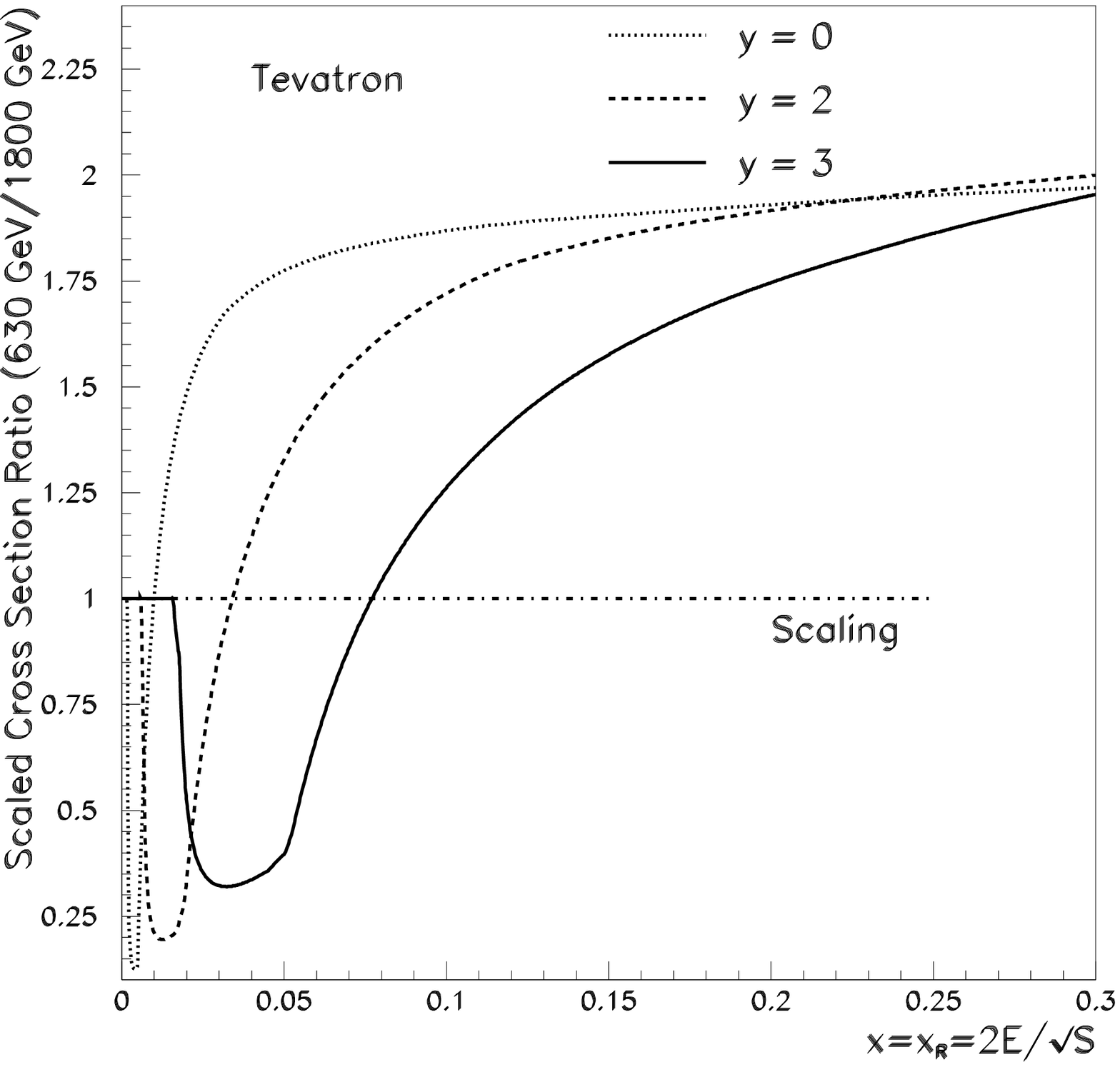}
\caption{"Dips" at Tevatron energies  \hspace*{1.4cm}
Figure 2: "Dips" at LHC energies}
\end{figure}
\begin{figure}[htb]
\vskip 0 cm
\includegraphics{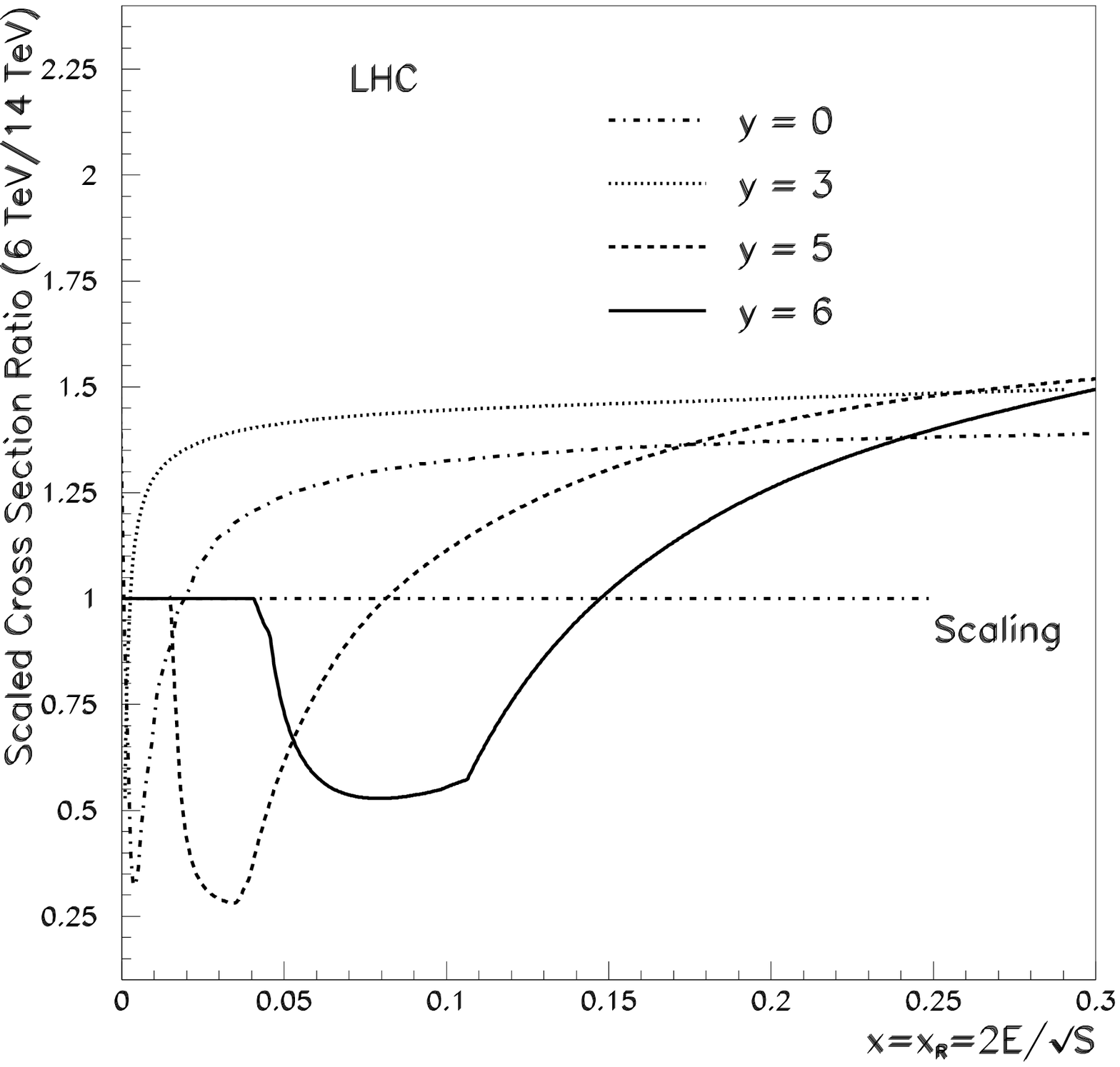}
\end{figure}

In the region of small $x$ (the only one we are interested in), 
the convolution $I(x,y,Q^2)$ of the parton distribution functions 
can be expressed (if one neglects powers of small $x$) 
via the effective parton distribution functions 
$F(x, Q^2)$ and the effective subprocess of Ref. \cite{Com84}:
\begin{equation}
\label{I}
I(x,y,Q^2)=\int^{z_{max}}_{z_{min}}dzF(\phi_+(x,y;z),Q^2)F(\phi_-(x,y;z),Q^2),
\end{equation}
where
\begin{equation}
\phi_\pm(x,y;z)=\frac{x(1+z^{\pm 1})}{1+e^{\mp2y}},
\end{equation}
and $z_{max}$, $z_{min}$ are determined by 
$\phi_+(x,y;z_{max})=1$, $\phi_-(x,y;z_{min})=1$.
In the calculation, $F(x,Q^2)$ of the proton was used which coincides 
with that of the antiproton. Thus, the plots of Figs. 1-3 correspond
to $pp$ or $p\bar{p}$ collision.
 
There is an important issue concerning the accuracy of present 
leading order (LO) calculation. The most important 
advantage of the scaled cross section ratio is 
that this is the ratio of two perturbative series 
with the same coefficients and with different scales 
in the running coupling. These scales are defined 
by the two initial collision energies at fixed 
scaling variable. One can show that the theoretical 
accuracy of the ratio in LO of perturbative QCD 
is at least not less than the accuracy of NLO 
calculations for absolute cross sections.

\addtocounter{figure}{+1}
\begin{figure}[htb]
\vskip 7.5 cm
\includegraphics{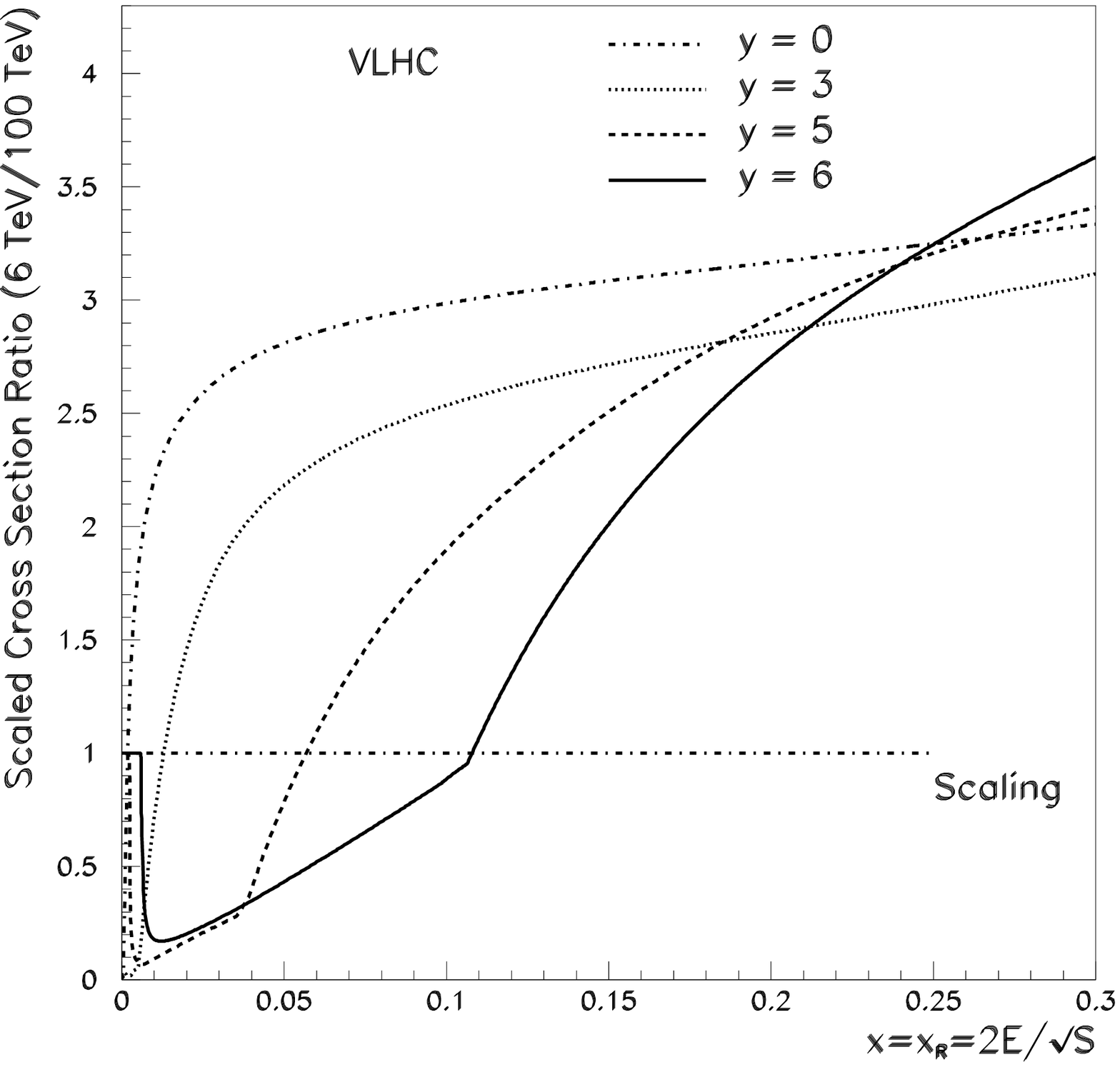}
\caption{"Dips" at VLHC energies}
\end{figure}

The minima in Figs. 1-3 originate from a competition between the 
running of the parton distribution functions and the running of the 
coupling constant. Namely, the ratio with frozen parton 
distribution functions is  decreasing monotonously (this tendency is
realized at small $x$),  while the one with frozen coupling  constant is
growing monotonously (this latter tendency is realized for
$x$ larger than the position of the minimum).

We suggest the following potential implications of the minima we
have predicted with the parton model: (i) If one observes 
the minima experimentally, one employs the orthodox parton model 
and tries to account for observed positions and depths  of the 
minima by taking into account higher order
corrections, in particular, resummation of the energy logarithms. 
(ii) If one does not observe the minima experimentally,
more radical changes are motivated such as an
alternative model of the elementary constituents 
inside the hadrons for semi-hard asymptotics. 
One example might be the color 
dipole model \cite{Mue94}.

Finally, we comment on the possibility for searching 
the minima at the Fermilab  Tevatron, at the CERN LHC, and at the VLHC: 
Positions of the minima for  the Tevatron energies (see Fig. 1)
seem to be reached 
for both D$\emptyset$ and CDF detectors. 
The minima of the LHC plot 
(see Fig. 2) seem to be well inside the acceptance of, e.g., 
the FELIX \cite{Felix}, the ALICE \cite{Alice} and 
the CMS \cite{CMS} detectors. 

We take the ratio of 6 TeV/14 TeV for the LHC, 
because, in addition to 14 TeV $pp$ collisions, 
lead-lead collisions at the LHC are planned with  
the collision energy  of 6 TeV per nucleon-nucleon collision. 
Since nuclear collisions bring in nuclear effects which can distort
our predicted curves, we also considered the ratio 6 TeV/100
TeV (Fig. 3).  These latter predictions correspond to the VLHC 
energies \cite{VLHC}. 

Further consideration should be given
for which pair of energies and  value of rapidity are  most
convenient for an experimental search of the minima.
Also, more work is needed to make quantitative predictions
for the locations and the shapes of the dips with the NLO
corrections taken into account.

It is worth noting that in the case of nuclear collisions the
effects of initial nuclear parton distributions (small-x EMC-effect
\cite{EMC}) and dynamical effects like quark-gluon plasma, 
jet quenching, {\it etc.} \cite{Wang97}
will demand special consideration. We plan to consider
these and related issues in future efforts.

Before making conclusions, we would like to note that many of
the above ideas can be studied also for the case of heavy quarkonium
production, where similar phenomena should be present
\cite{KPSV}.

To sum up, we find a new qualitative prediction of the 
QCD-improved parton model for hadron collisions and 
suggest its use to test the applicability of the 
parton model for certain regions of high energy
hadron collisions.

 We thank A. Akopian, D.S. Denisov, K. Eggert, J. Huston, A.L. Kataev, 
 J. Krane, V.A. Kuzmin, S.A. Larin, L.N. Lipatov, V.A. Matveev, 
 N.K. Terentyev, 
 N. Varelas,  A.A. Vorobyov, H. Weerts and P.I. Zarubin for 
 helpful discussions.
 VTK and GBP thank the International Institute of Theoretical
 and Applied Physics of the Iowa State University for their warm 
 hospitality and support. We thank the National Energy Research
 Scientific Computing Center (NERSC) at the LBNL for 
 high-performance computing resources.
 This work was supported in part by the Russian Foundation
 for Basic Research, grant Nos. 96-02-16717 and 96-02-18897 
 and by the U.S. Department of Energy, grant No. 
 DE-FG02-87ER40371, Division of High Energy and
 Nuclear Physics.


\begin{thebibliography}{10}

\bibitem{Iof84}
B.L. Ioffe, V.A. Khoze and L.N. Lipatov, 
{\it Hard Processes. Vol.1: Phenomenology, Quark Parton Model} 
(North Holland, Amsterdam, 1984); \\
R.D. Field, {\it Applications of Perturbative QCD}
(Addison-Wesley, Redwood City, 1989)
Frontiers in Physics, Vol. 77

\bibitem{CTEQ95}
CTEQ Collaboration, R. Brock {\it et al.},
Rev. Mod. Phys. {\bf 67}, 157 (1995); \\
H.L. Lai {\it et al.}, Phys. Rev. {\bf D55}, 1280 (1997)

\bibitem{Col82}
P.D.B. Collins, 
{\it An Introduction to Regge Theory and High-Energy Physics} 
(Cambridge, England, 1977)

\bibitem{Lip76}
 L.N. Lipatov, Yad. Fiz. {\bf 23}, 642 (1976)
 [Sov. J. Nucl. Phys. {\bf 23}, 338 (1976)]; \\
 E.A. Kuraev, L.N. Lipatov and V.S. Fadin, Zh. Eksp. Teor. Fiz.
 {\bf 71}, 840 (1976) [Sov. JETP {\bf 44}, 443 (1976)];
 {\bf 72}, 377 (1977) [{\bf 45}, 199 (1977)]; \\
 Ya.Ya. Balitski\v i and L.N. Lipatov, Yad. Fiz. {\bf 28}, 1597 (1978)
 [Sov. J. Nucl. Phys. {\bf 28}, 822 (1978)];  \\
 L.N. Lipatov, Zh. Eksp. Teor. Fiz. {\bf 90}, 1536 (1986)
 [Sov. JETP {\bf 63}, 904 (1986)]

\bibitem{Lip97}
L.N. Lipatov, Phys. Rep. {\bf C286}, 131 (1997)

\bibitem{Bal97}
R.D. Ball and S. Forte, Phys. Lett. {\bf 405B}, 317 (1997)

\bibitem{Ynd96}
F. Barreiro, C. L\'opez and F.J. Yndur\'ain,
Zeit. Phys. {\bf C72}, 561 (1996); \\
F.J. Yndur\'ain, FTUAM 96-19 (1996) Madrid, hep-ph/9605265


\bibitem{HERA98}
ZEUS Collaboration, J. Breitweg
{\it et al.}, DESY-98-050 (1998), 
hep-ex/9805016 \\
H1 Collaboration, C. Adloff {\it et al.}, DESY-98-143 (1998), 
hep-ex/9809028, submitted to Nucl. Phys. 


\bibitem{Bar96}
J. Bartels, V. Del Duca, A. De Roeck, D. Graudenz, 
M. W\"usthoff, Phys. Lett. {\bf 384B}, 300 (1996) 


\bibitem{UA285}
UA2 Collaboration, J.A. Appel {\it et al.}, 
Phys. Lett. {\bf 160B}, 349 (1985); \\
UA1 Collaboration, G. Arnison {\it et al.}, 
Phys. Lett. {\bf 172B}, 461 (1986); \\
CDF Collaboration, F. Abe {\it et al.}, 
Phys. Rev. Lett. {\bf 62}, 613 (1989)


\bibitem{CDF93}
CDF Collaboration, F. Abe {\it et al.}, Phys. Rev. Lett. {\bf 70}, 1376 (1993)
 

\bibitem{CDF96}
CDF Collaboration, T. Devlin, {\it Proc. of XXVIII ICHEP-96},
July 25-31, 1996, Warsaw, Poland, edited by  Z. Ajduk and
A.K. Wroblewski (World Scientific, Singapore, 1997); \\
A.A. Bhatti, {\it Proc. of the DPF96 Meeting}, August 10-15,
1996, Minneapolis, MN (World Scientific, Singapore, 1997)


\bibitem{Hus98}
J. Huston, Plenary talk at the {\it XXIX Int. Conf. on High Energy Physics
-- ICHEP-98},  Vancouver, Canada, July 23 - 29, 1998 
(World Scientific, Singapore), hep-ph/9901352; \\
CDF Collaboration (preliminary), A. Akopian, PhD thesis, Rockefeller Univ. (1999), \\ 
$ http://www-cdf.fnal.gov/physics/new/qcd/QCD.html $; \\
D0 Collaboration (preliminary), J. Krane, PhD thesis, Nebraska Univ. (1999), \\ 
$ http://www-d0.fnal.gov/results/publications\_talks/thesis/thesis.html $

\bibitem{Ell93}
S.D. Ellis, {\it Proc. of the XXVIIIth Rencontres de Moriond},
Les Arcs, France, March 20-27, 1993, edited by J. Tran Thanh Van
(Editions Frontieres, 1993) Vol.2, p.235

\bibitem{Gie96}
W. Giele, talk presented at the {\it CTEQ Symposium, Confronting 
QCD with Experiment: Puzzles and Challenges}, Fermilab, IL, 
November 7-9, 1996

\bibitem{Ell90}
S.D. Ellis, Z. Kunszt and D.E. Soper, Phys. Rev. Lett. {\bf 64},
2121 (1990); \\
F. Aversa, M. Greco, P. Chiappetta and J.Ph. Guillet,
Phys. Rev. Lett. {\bf 65}, 401 (1990); \\
W. Giele, E.W.N. Glover and D.A. Kosower, Phys. Rev. Lett. {\bf 73},
2019 (1994)

\bibitem{KP96b}
  V.T. Kim and G.B. Pivovarov, Phys. Rev. {\bf D57}, R1341 (1998)

\bibitem{VLHC} 
C.W. Foster and E. Malamud, {\it Low Cost Hadron Colliders at Fermilab: 
 a Discussion Paper}, FERMILAB-TM-1976 (1996); \\ 
 Talks in {\it Proc. of the DPF/DPB Summer Study on New
 Directions for High-Energy Physics, Snowmass '96}; \\
 Talks at {\it the Very Large Hadron Collider Physics and Detector
 Workshop}, Fermilab, March 13-15, 1997

\bibitem{CDF98}
CDF Collaboration, F. Abe {\it et al.}, Phys. Rev. {\bf D57}, 67 (1998)

\bibitem{Com84}
B.L. Combridge and C.J. Maxwell, Nucl. Phys. {\bf B239}, 429 (1984)

\bibitem{Mue94}
 A.H. Mueller, Nucl. Phys. {\bf B415}, 373 (1994); \\
 N.N. Nikolaev and B.G. Zakharov, Phys. Lett. {\bf 327B}, 149 (1994)


\bibitem{Felix}
 K. Eggert and C. Taylor, {\it FELIX: a Full Acceptance Detector for
 the CERN LHC}, CERN-PPE-96-136 (1996); \\
 FELIX Collaboration, E. Lippmaa {\it et al.}, 
 Letter of Intent, CERN/LHCC 97-45 (1997)
 

\bibitem{Alice}
 ALICE Collaboration, {\it Technical Proposal}, CERN/LHCC 95-71 (1995)
 
 \bibitem{CMS}
 CMS Collaboration, {\it Technical Proposal}, CERN/LHCC 94-38 (1994)
  
\bibitem{EMC}
EMC Collaboration, J.J. Aubert {\it et al.}, 
Nucl. Phys. {\bf B293}, 740 (1987); \\
M. Arneodo, Phys. Rept. {\bf C240}, 301 (1994)

\bibitem{Wang97}
X.N. Wang, Phys. Rept. {\bf C280}, 287 (1997)

\bibitem{KPSV}
V.T. Kim, G.B. Pivovarov, H.S. Song and J.P. Vary, in progress 

\end{thebibliography}
\end{document}